\documentclass[preprint,3p,times,authoryear]{elsarticle}

\usepackage{amssymb}
\usepackage{amsmath}
\usepackage{booktabs}

\usepackage[modulo]{lineno}

\usepackage[x11names]{xcolor}

\usepackage{tabularx}
\usepackage{cancel}

\usepackage{fancyref}
\usepackage{hyperref}
\hypersetup{colorlinks,breaklinks}
\usepackage{xspace}

\newcommand{\dedalus}{{\sc{Dedalus}}\xspace}
\newcommand{\pytorch}{{\texttt{PyTorch}}\xspace}
\newcommand{\pinninit}{{\texttt{pinninit}}\xspace}
\newcommand{\pinninitrepo}{\url{https://github.com/tdsandnes/pinninit}}
\newcommand{\coderepo}{\url{https://github.com/tdsandnes/pinn-1d-mhd-validation}}

\newcommand\nonumfootnote[1]{%
  \begingroup
  \renewcommand\thefootnote{}\footnote{#1}%
  \addtocounter{footnote}{-1}%
  \endgroup
}

\journal{Journal of Computational Physics}

\begin{document}

\begin{frontmatter}



\title{Validation of Physics-Informed Neural Networks in a One-Dimensional Magnetohydrodynamic Inverse Problem} 


\author[Leeds]{T.~D.~Sandnes$^{*,\,}$}
\author[Guilin]{L.~Chen}
\author[Leeds]{D.~P.~Cummins}
\author[Leeds]{C.~J.~Davies}
\author[Macau]{J.~Li}
\author[Leeds]{P.~W.~Livermore}

\affiliation[Leeds]{organization={School of Earth, Environment and Sustainability, University of Leeds},
             addressline={Woodhouse},
             city={Leeds},
             postcode={LS2 9JT},
             country={UK}}
\affiliation[Guilin]{organization={School of Earth Sciences, Guilin University of Technology},
             city={Guilin},
             postcode={541004},
             country={China}}
\affiliation[Macau]{organization={Macau Institute of Space Technology and Application, Macau University of Science and Technology},
             city={Macao},
             postcode={999078},
             country={China}}
             
\nonumfootnote{
\footnotesize{
$^*$Corresponding author, 
\href{mailto:t.d.sandnes@leeds.ac.uk}{t.d.sandnes@leeds.ac.uk}.}}

\begin{abstract}
Physics-informed neural networks (PINNs) provide a means of solving inverse problems by combining incomplete observations with governing equations to reconstruct the physical state of a system. We test this approach using a one-dimensional system of magnetohydrodynamic (MHD) Burgers' equations, in which counter-propagating Alfv\'enic fluctuations, driven at one boundary and reflected at the other, interact through nonlinear coupling between the velocity and magnetic fields to produce a continually evolving state. Using magnetic field measurements at a single boundary as the only time-dependent observations, we find that a basic PINN formulation fails to recover the interior dynamics. However, both the velocity and magnetic fields can be accurately reconstructed throughout the domain using a combination of methodological improvements: the SOAP optimiser, a SIREN architecture, dynamic loss-term weighting, and residual-based attention. Each method contributes cumulatively to reconstruction accuracy, with the higher-order optimiser playing a key role. Reconstruction errors increase with the number of driving frequencies and are lowest near the observed boundary. To test reconstruction with fewer constraints, we omit the boundary-condition loss term on the magnetic field gradient, finding that independently trained PINNs consistently converge towards closely matching, low-loss solutions that differ from the true system state. This highlights that repeatable PINN convergence and low loss values do not necessarily indicate successful recovery of the underlying system when training is under-constrained. This work demonstrates the potential of PINNs for reconstructing unobserved interior dynamics in MHD from boundary observations and provides a controlled testbed for validating and benchmarking methods for inverse problems.
\end{abstract}

\begin{keyword}
Physics-informed neural networks \sep
Magnetohydrodynamics \sep
Inverse problems


\end{keyword}

\end{frontmatter}


\section{Introduction}\label{sec:intro}

Inverse problems, in which the state or properties of a system are to be inferred from incomplete observations, arise across a wide range of applications in science and engineering \citep{tarantola2005inverse}.
Available observations often provide only indirect or partial information about the quantities of interest, which themselves may be inaccessible to direct measurement.
Solving such problems therefore requires combining available data with additional constraints, such as physical models that describe the underlying system.
This process becomes particularly challenging for nonlinear, time-dependent systems, where complex dynamics can make the inverse problem ill-conditioned and computationally expensive.

Magnetohydrodynamics (MHD) describes the coupled evolution of electrically conducting fluids and magnetic fields, providing a framework for modelling a wide range of astrophysical and geophysical flows, as well as laboratory plasmas \citep{chandrasekhar1960hydrodynamic}.
In many MHD systems, observations are available predominantly at or outside the system boundary, while the interior dynamics cannot be measured directly, leading to challenging inverse problems.
In geomagnetism, for example, measurements of Earth's time-varying magnetic field provide constraints on the field at the core--mantle boundary, but only indirect information about the flow and the deeper magnetic, thermal, and compositional structure of the liquid outer core \citep{fournier2010introduction}.
Similarly, in tokamaks, measurements made primarily outside the plasma are used to infer the internal equilibrium, including the magnetic field, current density, and pressure \citep{rossi2023potential}.
To evaluate the capabilities of numerical methods without the full complexity of realistic systems, simplified one-dimensional MHD models \citep{thomas1968numerical} provide useful testbeds for investigating inverse problems, and have previously been used to study data assimilation in MHD systems \citep{fournier2007case, sun2007data}.

The rapid development of scientific machine learning has produced a diverse range of approaches for combining observational data with physical models \citep{karniadakis2021physics, dietrich2025scientific}.
These include neural operators, which learn mappings between function spaces \citep{lu2019deeponet, li2020fourier}; equation-discovery methods, which infer governing equations directly from data \citep{brunton2016discovering}; and differentiable solvers, which enable optimisation of numerical simulations using gradients propagated through the solver \citep{newbury2024review}.
Machine learning approaches are increasingly being applied to MHD systems \citep{rossi2023potential, li2025reconstruction}, alongside the development of dedicated benchmarks for systematic comparison of different methods \citep[e.g.][]{rousseau2026tokamark}.

Physics-informed neural networks (PINNs) are a class of scientific machine learning methods that combine governing partial differential equations (PDEs) and observational data into a loss function that is minimised during training \citep{raissi2019physics}.
Earlier work pioneered neural network-based approaches for solving PDEs \citep{dissanayake1994neural, lagaris1998artificial}, while automatic differentiation helped enable the modern PINN formulation \citep{baydin2018automatic}.
PINNs have been applied across fluid dynamics \citep{raissi2020hidden, cai2021physics}, biomedicine \citep{sarabian2022physics}, geophysics \citep{rasht2022physics, li2025core, li2025reconstruction, shakespeare2025local, styp2025physics}, planetary science \citep{livermore2024reconstructions, delcourt2025modeling}, and many other fields \citep{toscano2025pinns}.

However, PINN training presents several challenges that can introduce errors into the resulting solution.
These include spectral bias, where networks preferentially learn low-frequency structure and struggle to represent fine-scale features \citep{rahaman2019spectral, khodakarami2026spectral}; unbalanced loss contributions, where competing objectives can lead to stiff and conflicting gradient dynamics \citep{wang2021understanding, wang2025gradient}; and propagation failure, where networks converge to incorrect or trivial solutions when information fails to propagate effectively throughout the domain \citep{daw2022mitigating}.
Numerous modifications to the basic PINN formulation have been developed to mitigate these issues.
In this work, we combine several such methods: a SIREN architecture to improve the representation of high-frequency features \citep{sitzmann2020implicit}, dynamic loss-term weighting to balance the contributions of the loss terms \citep{wang2021understanding, wang2023expert}, residual-based attention to adaptively weight points within each loss term \citep{anagnostopoulos2024residual}, and the SOAP optimiser to improve optimisation in the presence of conflicting gradients \citep{vyas2025soap, wang2025gradient}.

Existing applications of PINNs to MHD inverse problems have made use of simplifying assumptions, such as a steady state \citep{rossi2023potential} or frozen flux \citep{li2025reconstruction}, while reconstructions based on the full time-dependent MHD equations have used observations sampled within the interior of the domain \citep{bard2021neural, isayama2026reconstruction}.
The PINN methods described above have primarily been tested on forward problems, either individually or in partial combinations \citep{wang2023expert, wang2025gradient, khodakarami2026spectral}.
However, how these methods interact when combined in inverse problems where the interior state is constrained only indirectly by boundary observations, and how each contributes to reconstruction accuracy, remain to be established.

Here we validate the use of PINNs for recovering interior flow from boundary magnetic field measurements in a one-dimensional system of coupled MHD equations.
In \S\ref{sec:1D_MHD_system} we introduce the one-dimensional MHD system.
In \S\ref{sec:PINNs} we describe our PINN formulation, including the network architecture, loss function, sampling strategy, optimiser, and model-selection procedure.
In \S\ref{sec:network_and_samples} we present convergence tests motivating our choices of network size and numbers of training points.
In \S\ref{sec:results} we assess the PINN reconstruction of the system and investigate the sensitivity of our results to choices made in the PINN formulation.
We summarise our findings in \S\ref{sec:conclusions}.

\section{One-dimensional magnetohydrodynamic system}\label{sec:1D_MHD_system}

We consider a one-dimensional system of coupled MHD equations that serves as a testbed for investigating the recovery of interior flow from boundary magnetic field measurements.
We carry out simulations of the system to generate reference solutions, sample the magnetic field at the system boundary, and use PINNs to reconstruct the interior velocity and magnetic fields from these measurements, assessing the accuracy of reconstructions against the reference solutions.
The objective is to construct a system that retains several qualitative features of sustained three-dimensional MHD systems: the system should remain in a dynamically steady but non-repeating state, with fluctuations of broadly persistent amplitude and frequency; should maintain continual nonlinear coupling between the magnetic and velocity fields; and should avoid shock formation.

Burgers' equation is the simplest model that combines advection with viscous dissipation \citep{bateman1915some, burgers1948mathematical}. This model was extended to magnetohydrodynamics by \citet{thomas1968numerical} in order to study the turbulent dynamo problem, producing a one-dimensional analogue of the momentum and induction equations,
\begin{align}
    \frac{\partial u}{\partial t} + u \frac{\partial u}{\partial x} - B \frac{\partial B}{\partial x} &= \nu \frac{\partial^2 u}{\partial x^2} \;, \label{eq:thomas_eqns_u}\\
    \frac{\partial B}{\partial t} + u \frac{\partial B}{\partial x} - B \frac{\partial u}{\partial x} &= \eta \frac{\partial^2 B}{\partial x^2} \;, \label{eq:thomas_eqns_b}
\end{align}
where $u(x,t)$ is the velocity field, $B(x,t)$ is the magnetic field rescaled in velocity units, and $\nu$ and $\eta$ are the kinematic viscosity and magnetic diffusivity respectively.
Each field is advected by the flow, diffused, and nonlinearly coupled to the other.

In one dimension, imposing $\nabla \cdot \mathbf{B} \equiv \partial_x B = 0$ would imply a spatially uniform magnetic field, and imposing incompressibility, $\nabla \cdot \mathbf{u} \equiv \partial_x u = 0$, would likewise imply a uniform velocity field.
Neither constraint is therefore imposed in Eqs.~\ref{eq:thomas_eqns_u} and~\ref{eq:thomas_eqns_b}.
Without incompressibility, however, advection-dominated dynamics can lead to shock formation through the steepening of negative velocity gradients.
Although viscous dissipation can smooth this steepening, sufficiently strong dissipation also suppresses fluctuations and weakens the sustained dynamics that we aim to study.
We therefore consider a regime dominated by Alfv\'enic fluctuations propagating on a constant background magnetic field, writing $B = U_A + b$, where $U_A$ is the Alfv\'en speed associated with the constant background field, with magnetic fields expressed in velocity units and $b(x,t)$ denoting the evolving fluctuations.
The propagation of these fluctuations through the domain sustains nonlinear interactions while mitigating the gradient steepening associated with purely advective dynamics, without requiring strong dissipation that would suppress small-scale structure.

We non-dimensionalise lengths by the domain size $L$, fluctuation amplitudes by a velocity scale $U$ (so that $u \sim b \sim U$), and time by the Alfv\'en crossing time $T \equiv L/U_A$. The dynamics are then governed by three dimensionless numbers: the Alfv\'en Mach number, $M_A \equiv U/U_A$, the Reynolds number, $Re \equiv UL/\nu$, and the magnetic Reynolds number, $R_m \equiv UL/\eta$.
The resulting system can be expressed in terms of the Els\"asser variables $z^\pm \equiv u \pm b$ \citep{elsasser1950hydromagnetic}, using the dimensionless velocity and magnetic fluctuations. The equations become
\begin{equation}\label{eq:elsasser_eqns}
    \frac{\partial z^\pm}{\partial t} \mp \frac{\partial z^\pm}{\partial x} + M_A \, z^\mp \frac{\partial z^\pm}{\partial x} = \nu^+ \frac{\partial^2 z^\pm}{\partial x^2} + \nu^- \frac{\partial^2 z^\mp}{\partial x^2} \;,
\end{equation}
where $\nu^\pm \equiv M_A \left(1/Re \pm 1/R_m\right)/2$ (see \ref{app:derivation} for details of the derivation).
In the absence of the nonlinear and diffusive terms, $u$ and $b$ each satisfy a wave equation, for which $z^+$ and $z^-$ are the characteristic variables.
The Els\"asser variables therefore represent counter-propagating Alfv\'enic fluctuations, with $z^+$ propagating at unit speed in the negative-$x$ direction and $z^-$ in the positive-$x$ direction.
The nonlinear term, $M_A z^\mp \partial_x z^\pm$, couples each of $z^\pm$ to the oppositely-propagating field $z^\mp$.
The diffusive terms consist of self-diffusion of each field, with coefficient $\nu^+$, and a cross-diffusion between the two fields, with coefficient $\nu^-$. The latter vanishes if $Re = R_m$.
In our simulations, we evolve $z^\pm$ by solving Eq.~\ref{eq:elsasser_eqns} directly; the velocity and magnetic field fluctuations are then recovered as $u = (z^+ + z^-)/2$ and $b = (z^+ - z^-)/2$.

\subsection{Fiducial system}\label{subsec:fiducial}

\begin{figure}
	\centering
{\includegraphics{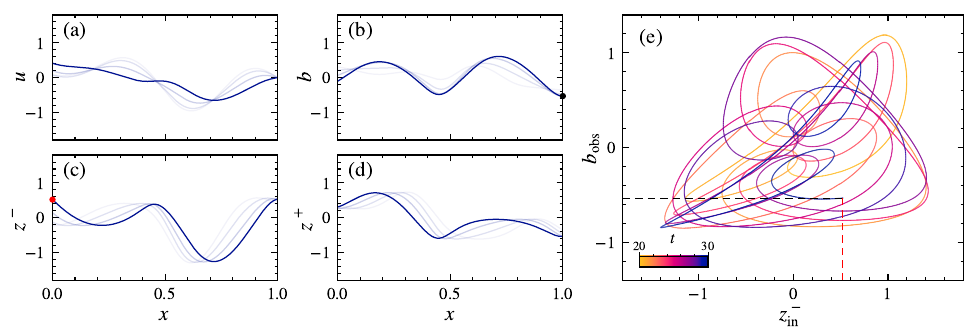}
}
    \vspace{-0.25in}
	\caption{Two-frequency fiducial system at $t=30$: (a) velocity, $u$; (b) magnetic field perturbation, $b$; (c) rightward-propagating $z^- = u - b$; and (d) leftward-propagating $z^+ = u + b$. Three earlier profiles are included at intervals of $0.05$ before $t=30$, plotted progressively lighter in colour with earlier time. Panel (e) shows the trajectory in $b_{\rm obs}$--$z^-_{\rm in}$ phase space over the interval $t \in [20,30]$, coloured by $t$. The corresponding boundary field values at $t=30$ are indicated by black and red markers in (b) and (c), and by dashed lines of the same colours in (e). The system has dimensionless numbers $Re = R_m = 500$ and $M_A = 1$.}
	\label{fig:snapshot_attractor_2freq_1Ma_500Re_500Rm}
\end{figure}

We consider the system of Eq.~\ref{eq:elsasser_eqns} on the spatial domain $x \in [0, 1]$.
At the left boundary, $x = 0$, we force the system through the rightward-propagating variable $z^-$, injecting fluctuations that propagate into the domain.
At this boundary, the leftward-propagating variable $z^+$ is given a Neumann condition so as not to over-constrain the fluctuations leaving the domain.
At the right boundary, $x = 1$, we impose a non-penetration condition, $u = 0$, on the velocity together with a Neumann condition on $b$, which leaves its boundary value free to vary as the system evolves.
The non-penetration condition reflects the rightward-propagating fluctuations $z^-$ back into the domain as leftward-propagating $z^+$, which then interact with newly driven waves entering at $x=0$.
The boundary conditions are summarised as
\begin{equation}\label{eq:fiducial_bcs}
\begin{alignedat}{2}
    z^-(0, t) &= z^-_{\rm in}(t) \;, &\qquad u(1, t) &\equiv \tfrac{1}{2}\left[z^+(1,t) + z^-(1,t)\right] = 0 \;, \\
    \partial_x z^+(0, t) &= 0 \;, &\qquad \partial_x b(1, t) &\equiv \tfrac{1}{2}\left[\partial_x z^+(1,t) - \partial_x z^-(1,t)\right] = 0 \;.
\end{alignedat}
\end{equation}
Since $u$ and $b$ are linear combinations of the evolved variables $z^\pm$, the boundary conditions at $x=1$ are imposed in the simulation as the equivalent conditions on $z^\pm$ and their derivatives.
For the training of PINNs in \S\ref{sec:network_and_samples} and \S\ref{sec:results}, magnetic field measurements are taken only at the $x=1$ boundary, with $b_{\rm obs}(t) \equiv b(1,t)$ denoting the observed boundary magnetic field.

We take the forcing to be the sum of $N_f$ sinusoids of equal amplitude,
\begin{equation}\label{eq:forcing}
    z^-_{\rm in}(t) = \frac{1}{\sqrt{N_f}} \sum_{i=1}^{N_f} \sin\!\left(2\pi f_i t\right) \;,
\end{equation}
where the amplitude is scaled by the factor $1/\sqrt{N_f}$ to fix the total forcing power across cases with different numbers of components.
The frequencies $f_i$ are chosen to be irrational and to not be rational multiples of one another, so that no forcing component shares a common period with the Alfv\'en crossing time or with any other component.
Together with the nonlinear interaction of the counter-propagating waves, this aims to prevent the system from settling into a trivial periodic cycle.
We consider three forcing frequencies, $f_1 = 1/\sqrt{2}$, $f_2 = \sqrt{3}$, and $f_3 = \sqrt{5}$, and investigate cases with one, two, and three frequencies, corresponding to forcing at $\{f_1\}$, $\{f_1, f_2\}$, and $\{f_1, f_2, f_3\}$, respectively.
Increasing the number of forcing frequencies by adding progressively higher-frequency components enables PINN performance to be evaluated across a controlled range of system complexities, with each additional component introducing finer-scale structure into the interior dynamics.

We take $Re = R_m = 500$ and $M_A = 1$ as our fiducial dimensionless numbers. 
Equal Reynolds and magnetic Reynolds numbers give $\nu^- = 0$, so the cross-diffusion term in Eq.~\ref{eq:elsasser_eqns} vanishes, simplifying the dissipation behaviour.
We choose these Reynolds numbers to be large enough that dissipation does not significantly damp fluctuations on the scales of interest, but small enough to avoid sharp gradients forming near the fixed $u = 0$ boundary at $x = 1$.
The choice of $M_A = 1$ ensures that nonlinear interaction and wave propagation contribute comparably to the dynamics.

The forward model is solved using the open-source spectral solver \dedalus \citep{burns2020dedalus}.
The evolved variables $z^\pm$ are represented using a Chebyshev basis with $256$ modes and a dealiasing factor of $3/2$.
Boundary conditions are enforced using the generalised tau method \citep{burns2020dedalus}.
The equations are integrated in time with the SBDF2 time stepper \citep{wang2008variable} and a fixed time step $\Delta t = 1 / 1024$.
We take smaller time steps than strictly required for the stability or accuracy of the solver, so as to densely sample the measurements needed for training the PINNs without need for interpolation.
Our choice of $\Delta t$ gives $2^{10}+1$ samples per unit time, including both endpoints, allowing lower temporal sampling densities to be generated by successive factors of two without interpolation.
The system is initialised from rest with no magnetic perturbation, $z^\pm(x, 0) = 0$, and simulated to $t = 30$.
To exclude the transient adjustment from the initial state, all PINNs are trained on $t \in [20,30]$, a span of ten Alfv\'en crossing times that we refer to as the training window.
Field values are sampled on a uniform grid of $257$ points in $x$ (spacing $1/256$, including both endpoints), evaluated directly from the spectral representation.
The boundary measurements and the test points are drawn from this grid so no spatial interpolation is required to construct them.

\begin{figure}
	\centering
{\includegraphics{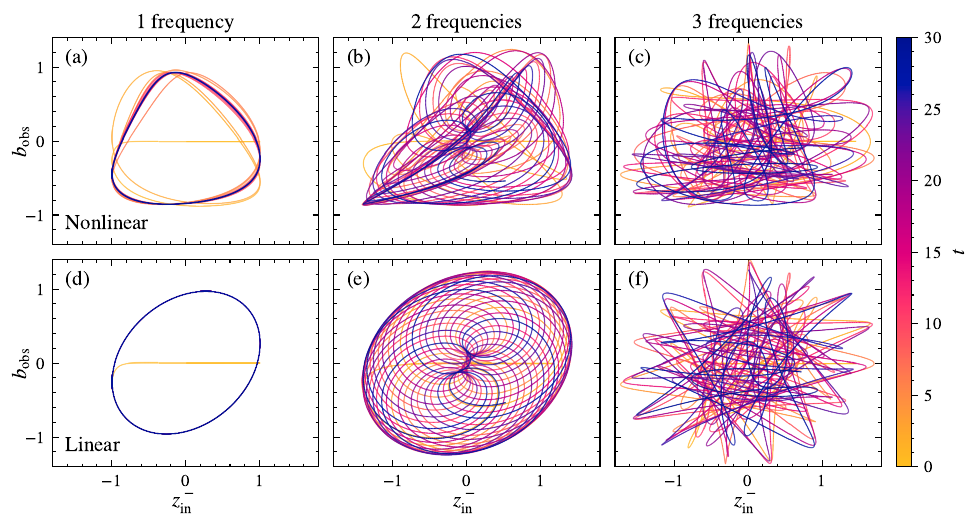}
}
    \vspace{-0.25in}
	\caption{Trajectory in $b_{\rm obs}$--$z^-_{\rm in}$ phase space of the fiducial system, for (a, d) one; (b, e) two; and (c, f) three forcing frequencies. Systems shown in (a--c) include the nonlinear term, $M_A z^\mp \partial_x z^\pm$, in the governing equations, Eq.~\ref{eq:elsasser_eqns}, while those in (d--f) omit it. Trajectories are shown over the interval $t \in [0,30]$ and coloured by $t$. Systems have dimensionless numbers $Re = R_m = 500$ and $M_A = 1$.}
	\label{fig:attractor_grid}
\end{figure}

A snapshot of the two-frequency fiducial system at $t=30$ is shown in Fig.~\ref{fig:snapshot_attractor_2freq_1Ma_500Re_500Rm}(a--d).
The fluctuations of $z^+$ are of comparable magnitude to those of $z^-$.
This behaviour is observed across all three forcing-frequency cases, and in additional tests with the nonlinear term disabled, confirming that diffusion does not significantly damp the primary fluctuations over the time taken for them to traverse the domain, reflect, and propagate back.
Over the training window, we compare the measured $b_{\rm obs}$ with the imposed $z^-_{\rm in}$ at each time.
As shown in Fig.~\ref{fig:snapshot_attractor_2freq_1Ma_500Re_500Rm}(e), the resulting trajectory does not collapse onto an orbit or cycle, consistent with the desired non-repeating behaviour.

Similar system trajectories for $t \in [0, 30]$ are shown in Fig.~\ref{fig:attractor_grid} for all three forcing-frequency cases, as well as for cases where the nonlinear term in Eq.~\ref{eq:elsasser_eqns}, $M_A z^\mp \partial_x z^\pm$, is set to zero.
Although these visualisations consider fields at only two positions rather than the full system, they show that the dynamical complexity increases as higher-frequency components are added to the forcing, and that the nonlinear term increases it further.
In the single-frequency case, the system nevertheless finds a repeating orbit, though one distinct from the trivial periodicity of the linear case.

Code for running the simulations and sampling the training, validation, and test sets is publicly available.\footnote{\label{fn:code}\coderepo. v1.0.0 was used for all results presented here.}

\section{Physics-informed neural networks}\label{sec:PINNs}

A neural network is a mapping $\mathcal{N}_{\boldsymbol{\theta}} : \mathbb{R}^{N_{\rm in}} \to \mathbb{R}^{N_{\rm out}}$; $\mathbf{x} \mapsto \mathcal{N}_{\boldsymbol{\theta}}(\mathbf{x})$, from $N_{\rm in}$ inputs to $N_{\rm out}$ outputs, defined by its architecture (its functional form) and the model parameters $\boldsymbol{\theta}$ \citep{goodfellow2016deep}.
Network architecture can be chosen such that $\mathcal{N}_{\boldsymbol{\theta}}$ is differentiable with respect to $\boldsymbol{\theta}$, allowing derivatives of its output to be evaluated by automatic differentiation \citep{baydin2018automatic}.
These gradients can be used to iteratively update the network parameters in order to minimise a prescribed loss function, in a process known as training.

Physics-informed neural networks \citep[PINNs;][]{raissi2019physics} are neural networks trained by minimising a loss function comprising data-misfit and PDE residual terms, driving the network towards a solution consistent with both observations of a system and its governing physics.
Since the network is also differentiable with respect to its inputs $\mathbf{x}$, which in our case are $\mathbf{x} = (x,t)$, the derivatives required for the PDE residuals can be evaluated at arbitrary coordinates.
Although the data and PDE constraints are imposed only at discrete points, the network itself is a continuous, mesh-free representation of the solution.

All PINNs presented here were trained using \pinninit, our open-source framework for initialising and training PINNs, which allows different systems of equations, problem geometries, network architectures and training methods to be implemented and combined flexibly.\footnote{\pinninitrepo. v0.1.0 was used for all results presented here.}
\pinninit is built on \pytorch \citep{paszke2019pytorch}, and derivatives are evaluated using its automatic differentiation functionality \citep{paszke2017automatic}.
Computations were performed with 64-bit floating-point precision \citep{xu2025fp64} and each PINN was trained on a single NVIDIA H200 NVL GPU.
The hyperparameters, and other settings used for all PINNs presented here, are summarised in \ref{app:hyperparameters}.
The code used to train and analyse every PINN presented here is publicly available.\textsuperscript{\ref{fn:code}}

\subsection{Network architecture}\label{subsec:architecture}

The choice of network architecture determines the representational capacity of the PINN and can strongly influence its optimisation dynamics \citep{wang2021understanding, wang2024piratenets}.
The simplest neural network architecture, and the one most standard for PINNs, is the multilayer perceptron (MLP), structured as a series of layers, each consisting of a linear combination of its inputs offset by a bias, followed by an elementwise nonlinear activation function.
Given a network input $\mathbf{x}^0 \equiv \mathbf{x}$, layer $\ell$ maps its own input $\mathbf{x}^{\ell-1} \in \mathbb{R}^{N_{\ell-1}}$ to
\begin{equation}
    \mathbf{x}^{\ell} = \sigma\!\left(\mathbf{W}^{\ell} \mathbf{x}^{\ell-1} + \mathbf{b}^{\ell}\right) \in \mathbb{R}^{N_\ell} \;, \qquad \ell = 1, \dots, L-1 \;,
\end{equation}
where $\mathbf{W}^\ell \in \mathbb{R}^{N_\ell \times N_{\ell-1}}$ and $\mathbf{b}^\ell \in \mathbb{R}^{N_\ell}$ are the trainable weight matrix and bias vector of layer $\ell$, and the activation function $\sigma$ is applied to each component independently.
Each component of $\mathbf{x}^\ell$ is the output of a neuron, and the number of neurons, $N_\ell$, is referred to as the width of layer $\ell$.
For unconstrained target fields, the activation function is not applied in the final layer, giving
\begin{equation}
    \mathcal{N}_{\boldsymbol{\theta}}(\mathbf{x}) = \mathbf{W}^{L} \mathbf{x}^{L-1} + \mathbf{b}^{L} \;.
\end{equation}
The weights and biases of all layers together constitute the network parameters, $\boldsymbol{\theta} = \{\mathbf{W}^\ell, \mathbf{b}^\ell\}_{\ell=1}^{L}$.

MLPs are universal function approximators: given sufficient width, they can approximate any continuous function on a compact domain to arbitrary accuracy \citep{cybenko1989approximation, hornik1989multilayer}.
In practice, however, a trained network will differ from the target function through three sources of error \citep{lu2021deepxde}: approximation error, arising from limitations in the representational capacity of the network architecture; generalisation error, since minimising the loss at a finite set of training points does not necessarily ensure accurate predictions elsewhere in the domain; and optimisation error, since training may fail to find a global minimum of the loss function.
More fundamentally, a unique, well-defined target solution is not guaranteed for inverse problems, where the available observations may not uniquely determine the system state, so that a solution minimising the loss need not correspond to the true underlying state.

In this paper we use a sinusoidal representation network architecture \citep[SIREN;][]{sitzmann2020implicit} based on sine activation functions, a specialised weight initialisation scheme that enables faster and more reliable convergence, and a modified first layer to improve the representation of high-frequency signals.
The derivatives of a SIREN network retain a sinusoidal structure, allowing the derivatives required for PDE residuals to remain well represented by the network under repeated differentiation.
SIREN has been shown to reduce spectral bias in PINNs, with negligible additional computational cost, for a range of optimisation methods \citep{khodakarami2026spectral}.

For the hidden layers, $\ell = 2, \dots, L-1$, SIREN layers take the form $\mathbf{x}^{\ell} = \sin(\mathbf{W}^{\ell}\mathbf{x}^{\ell-1} + \mathbf{b}^{\ell})$, while the final layer is linear.
The weights of all layers $\ell \geq 2$ are initialised from the uniform distribution $\mathbf{W}^{\ell} \sim \mathcal{U}(-\sqrt{6/N_{\ell-1}}, \sqrt{6/N_{\ell-1}})$, where $N_{\ell-1}$ is the width of the previous layer.
These bounds keep the distribution of neuron outputs approximately preserved from layer to layer \citep{sitzmann2020implicit}.
The first layer takes the form $\mathbf{x}^{1} = \sin(\omega_0\mathbf{W}^{1}\mathbf{x}^{0} + \mathbf{b}^{1})$, where the constant factor $\omega_0$ increases the frequency of the sine so that it spans multiple periods over the normalised input domain.
Following the implementation of \citet{sitzmann2020implicit}, we use $\omega_0 = 30$ and initialise the first-layer weights as $\mathbf{W}^{1} \sim \mathcal{U}(-1/N_{\rm in}, 1/N_{\rm in})$.
This initialisation is not stated explicitly by \citet{sitzmann2020implicit}, but is used in their accompanying code.
All biases are initialised with $\mathbf{b}^{\ell} = \mathbf{0}$.

For comparison, we also consider PINNs that use a more conventional architecture: a standard MLP with hyperbolic tangent activation functions.
For these networks, weights are initialised using Xavier initialisation \citep{glorot2010understanding} and biases are initialised as zero.

All PINNs presented here are constructed to represent the one-dimensional MHD system described in \S\ref{sec:1D_MHD_system}.
The networks therefore take two input coordinates, $\mathbf{x}=(x,t)$, and produce two output fields, $\mathcal{N}_{\boldsymbol{\theta}}(x,t)=\left(\hat{u}_{\boldsymbol{\theta}}(x,t), \, b_{\boldsymbol{\theta}}(x,t)\right)$.
Input coordinates are linearly rescaled before being passed to the network, mapping both $x\in[0,1]$ and the training window to $[-1,1]$ \citep{sitzmann2020implicit}.
Rather than taking the velocity field directly as a network output, we enforce the boundary condition $u(1,t)=0$ as a hard constraint by transforming the network output with
\begin{equation}\label{eq:hard_bc}
    u_{\boldsymbol{\theta}}(x,t)= \left(1-x\right) \hat{u}_{\boldsymbol{\theta}}(x,t) \;,
\end{equation}
so that $u_{\boldsymbol{\theta}}$ vanishes exactly at $x=1$ for all $\hat{u}_{\boldsymbol{\theta}}$ \citep{lu2021physics}. The $u_{\boldsymbol{\theta}}$ and $b_{\boldsymbol{\theta}}$ fields are then the PINN representations of the velocity and magnetic field perturbations, respectively.

\subsection{Loss function and adaptive weighting}\label{subsec:loss}

To solve the inverse problem of reconstructing the system described in \S\ref{sec:1D_MHD_system} from boundary measurements, we construct a loss function consisting of a single data term, two physics terms, and a boundary-condition term.
Of the four boundary conditions in Eq.~\ref{eq:fiducial_bcs}, only those at $x=1$ are given to the PINN: $u(1,t)=0$ is imposed as a hard constraint (Eq.~\ref{eq:hard_bc}) and $\partial_x b(1,t)=0$ through the boundary-condition loss term.
The boundary conditions at $x=0$, including the forcing $z^-_{\rm in}(t)$, are deliberately omitted, since in the inverse problem the driving of the system is assumed to be unknown.
Without them, the governing equations and boundary conditions given to the PINN do not form a closed problem, and the missing information must instead be inferred by the PINN from the time evolution of $b_{\rm obs}$.

Each loss term, which we index $j \in \{\mathrm{d}, z^+, z^-, \mathrm{bc}\}$, is evaluated on its own set of $N_j$ training points, $\{\mathbf{x}_i^{j}\}_{i=1}^{N_j}$, on which corresponding residuals $\mathcal{R}_j(\boldsymbol{\theta}; \mathbf{x}_i^j)$ are computed.
The data term measures the misfit between the network output $b_{\boldsymbol{\theta}}$ and the observations $b_{\rm obs}$ at the boundary data points $\{\mathbf{x}_i^{\rm d}\}_{i=1}^{N_{\rm d}} \equiv \{(1,t_i^{\rm d})\}_{i=1}^{N_{\rm d}}$, with residual
\begin{equation}\label{eq:residual_data}
\mathcal{R}_{\rm d}(\boldsymbol{\theta};\mathbf{x}) = b_{\rm obs}(t) - b_{\boldsymbol{\theta}}(\mathbf{x}) \;.
\end{equation}

The physics terms penalise the residuals of Eq.~\ref{eq:elsasser_eqns} at two sets of collocation points $\{\mathbf{x}_i^{z^\pm}\}_{i=1}^{N_{z^\pm}} \equiv \{(x_i^{z^\pm},t_i^{z^\pm})\}_{i=1}^{N_{z^\pm}}$ that span the domain.
Writing the network representations of the Els\"asser variables as $z^\pm_{\boldsymbol{\theta}} \equiv u_{\boldsymbol{\theta}} \pm b_{\boldsymbol{\theta}}$, the corresponding residuals are
\begin{equation}\label{eq:residual_physics}
    \mathcal{R}_{z^\pm}(\boldsymbol{\theta};\mathbf{x}) =
    \frac{\partial z^\pm_{\boldsymbol{\theta}}}{\partial t}
    \mp \frac{\partial z^\pm_{\boldsymbol{\theta}}}{\partial x}
    + M_A \, z^\mp_{\boldsymbol{\theta}} \frac{\partial z^\pm_{\boldsymbol{\theta}}}{\partial x}
    - \nu^+ \frac{\partial^2 z^\pm_{\boldsymbol{\theta}}}{\partial x^2}
    - \nu^- \frac{\partial^2 z^\mp_{\boldsymbol{\theta}}}{\partial x^2} \;,
\end{equation}
where the derivatives are computed by automatic differentiation.
Equivalently, the physics loss could be constructed from the equations for $u$ and $b$, but we choose to use the $z^\pm$ formulation, since the symmetry of the two equations gives physics loss terms of the same form.
This aims to provide an elegant problem set-up that avoids introducing asymmetries that might contribute to imbalanced training dynamics.

The boundary-condition term penalises violations of the Neumann condition on $b$ at points $\{\mathbf{x}_i^{\rm bc}\}_{i=1}^{N_{\rm bc}} \equiv \{(1,t_i^{\rm bc})\}_{i=1}^{N_{\rm bc}}$, with residual
\begin{equation}\label{eq:residual_bc}
    \mathcal{R}_{\rm bc}(\boldsymbol{\theta};\mathbf{x}) = \frac{\partial b_{\boldsymbol{\theta}}}{\partial x}(\mathbf{x}) \;.
\end{equation}
We enforce this condition softly through a loss term rather than imposing it through a transformation of the network output, as is done for the Dirichlet condition on $u$ described in \S\ref{subsec:architecture}, since a constraint on the derivative is less easily imposed by construction.

The full loss function is given by
\begin{align}\label{eq:pinn_loss}
    \mathcal{L}(\boldsymbol{\theta}) &= w_{\rm d} \mathcal{L}_{\rm d}(\boldsymbol{\theta})
    + w_{z^+} \mathcal{L}_{z^+}(\boldsymbol{\theta})
    + w_{z^-} \mathcal{L}_{z^-}(\boldsymbol{\theta})
    + w_{\rm bc} \mathcal{L}_{\rm bc}(\boldsymbol{\theta}) \notag\\
    &= w_{\rm d}
    \left\langle
        \lambda_{i}^{\rm d}
        \left[ \mathcal{R}_{\rm d}(\boldsymbol{\theta}; \mathbf{x}_i^{\rm d}) \right]^2
    \right\rangle_i
    + w_{z^+}
    \left\langle
        \lambda_{i}^{z^+}
        \left[ \mathcal{R}_{z^+}(\boldsymbol{\theta}; \mathbf{x}_i^{z^+}) \right]^2
    \right\rangle_i
    + w_{z^-}
    \left\langle
        \lambda_{i}^{z^-}
        \left[ \mathcal{R}_{z^-}(\boldsymbol{\theta}; \mathbf{x}_i^{z^-})\right]^2
    \right\rangle_i
    + w_{\rm bc}
    \left\langle
        \lambda_{i}^{\rm bc}
        \left[ \mathcal{R}_{\rm bc}(\boldsymbol{\theta}; \mathbf{x}_i^{\rm bc}) \right]^2
    \right\rangle_i
    \;,
\end{align}
where $\langle \cdot \rangle_i$ denotes an averaging operation over the $N_j$ training points of each term.
The loss-term weights $w_j$ set the relative contribution of each loss term, while the pointwise weights $\lambda_i^j$ act within each term, weighting the contribution of individual data and collocation points.
Fixed values are most typically used for these, with the most common choice being $1$ for all $w_j$ and $\lambda_i^j$ \citep{raissi2019physics}.
However, competing training objectives can lead to imbalanced optimisation dynamics, making it beneficial and, in some cases, necessary to update the weights dynamically during the training process \citep{wang2021understanding}.

We use the dynamic loss-term weighting scheme of \citet{wang2023expert}, which scales the weights $w_j$ so that all loss terms are trained at a comparable rate.
The instantaneous weights at a given training iteration are calculated as
\begin{equation}\label{eq:wang_23_w_hat}
    \hat{w}_j = \frac{\sum_k \|\nabla_{\boldsymbol{\theta}}\mathcal{L}_k\|}
    {\|\nabla_{\boldsymbol{\theta}}\mathcal{L}_j\|} \;,
\end{equation}
where $j,k \in \{\mathrm{d},z^+,z^-,\mathrm{bc}\}$ index the loss terms, and the gradients are with respect to the network parameters.
The instantaneous weights are then used to update exponential moving averages,
\begin{equation}\label{eq:wang_23_w}
    w_j = \alpha w_j^{\rm old} + (1-\alpha) \hat{w}_j \;,
\end{equation}
where $w_j^{\rm old}$ is the value of $w_j$ from the previous update.
We take $\alpha = 0.9$ and initial values $w_j = 1$.
These moving averages are the weights used in Eq.~\ref{eq:pinn_loss}.
We update the weights every 1000 training iterations, as we find that too frequent updates lead to unstable training.

For the pointwise weights $\lambda_i^j$, we use residual-based attention \citep[RBA;][]{anagnostopoulos2024residual}, with a slight modification to more clearly separate its role in weighting points within each loss term from the role of $w_j$ in balancing the loss terms themselves.
This method assigns greater weight to points with sustained high residuals relative to other points within the same loss term, thereby prioritising poorly constrained regions.
For each loss term, the pointwise weights are updated every iteration according to the residual magnitude at each point,
\begin{equation}\label{eq:rba_lambda}
    \lambda_i^j = \left[\gamma \sqrt{\lambda_i^{j,\rm old}}
    + \eta^* \frac{\left|\mathcal{R}_j(\boldsymbol{\theta};\mathbf{x}_i^j)\right|}
    {\max_k \left|\mathcal{R}_j(\boldsymbol{\theta};\mathbf{x}_k^j)\right|}\right]^2 \;,
\end{equation}
where $\lambda_i^j$ is the RBA weight for point $i$ of loss term $j$, $\lambda_i^{j,\rm old}$ is the value of $\lambda_i^j$ from the previous iteration, and the maximum over $k$ is taken only over the training points of that loss term.
The power of $2$ and square root arise only from a difference in notation with \citet{anagnostopoulos2024residual}, who define their pointwise weights to appear within the squares in Eq.~\ref{eq:pinn_loss}.
Following \citet{anagnostopoulos2024residual}, we take $\gamma = 0.999$ and $\eta^* = 0.01$.
In \citet{anagnostopoulos2024residual}, a mean is used for the averaging operation in Eq.~\ref{eq:pinn_loss}.
Here, we use weighted averages with respect to the weights $\lambda_i^j$, defined for a pointwise quantity $q_i$ as $\langle \lambda_i^j q_i \rangle_i \equiv \sum_i \lambda_i^j q_i / \sum_i \lambda_i^j$, so that the loss-term weights $w_j$ retain their interpretation as the relative weighting between loss terms without interference from the overall scale of $\lambda_i^j$.
We initialise with $\lambda_i^j = 1$ for all points at the start of training, rather than $0$ as in \citet{anagnostopoulos2024residual}, so that the weighted averages in Eq.~\ref{eq:pinn_loss} are defined from the first iteration.

In practice, within each iteration, the loss (Eq.~\ref{eq:pinn_loss}) and its gradient are first evaluated using the current weights, after which the resulting residuals are used to update $\lambda_i^j$ and, every 1000 iterations, $w_j$.
The updated weights enter the loss only in the following iteration, so that the residuals used to update $w_j$ and $\lambda_i^j$ are based on the same set of network parameters, since the calculation of $w_j$ depends on $\lambda_i^j$ through the gradients of the loss terms.
The resulting one-iteration lag has little practical effect, since $w_j$ and $\lambda_i^j$ are both moving averages dominated by their history, and $w_j$ is additionally held fixed for 1000 iterations between updates.

In the method comparisons presented in \S\ref{subsec:method_comparison}, we also consider the use of RBA (Eq.~\ref{eq:rba_lambda}) without dynamic loss-term weighting (Eqs.~\ref{eq:wang_23_w_hat} and~\ref{eq:wang_23_w}), instead setting $w_j = 1$ for all loss terms.
In this case, each loss term is instead divided by its number of points, so the averaging operation in Eq.~\ref{eq:pinn_loss} becomes $\langle \lambda_i^j q_i \rangle_i \equiv \sum_i \lambda_i^j q_i / N_j$, as is done by \citet{anagnostopoulos2024residual}.

\subsection{Sampling strategy}\label{subsec:sampling}

\begin{table}
\centering
\caption{Network size and numbers of training points, $N_j$, and validation points, $N_j^{\rm val}$, for the data ($j = {\rm d}$), $z^\pm$ physics ($j = z^\pm$), and boundary-condition ($j = {\rm bc}$) loss terms. These values are used for all PINNs presented here, except in the convergence tests in \S\ref{sec:network_and_samples}, where the tested quantity is varied while the others are held fixed at the values listed.}\label{tab:fiducial_config}
\vspace{0.10in}
\begin{tabular}{lr}
\toprule
Quantity & Value \\
\midrule
Hidden layers               & 7     \\
Neurons per hidden layer    & 80    \\
$N_{\rm d}$                 & 820   \\
$N_{\rm d}^{\rm val}$       & 205   \\
$N_{z^\pm}$                 & 32768 \\
$N_{z^\pm}^{\rm val}$       & 8192  \\
$N_{\rm bc}$                & 16384 \\
$N_{\rm bc}^{\rm val}$      & 4096  \\
\bottomrule
\end{tabular}
\end{table}

For each loss term in Eq.~\ref{eq:pinn_loss}, we define a set of $N_j$ training points together with a corresponding set of $N_j^{\rm val}$ validation points.
The training sets are used to compute the training loss and update the network parameters during optimisation.
The loss terms are also evaluated on their corresponding validation sets at every training iteration, with these validation points kept separate from parameter updates.
The resulting validation losses are used to assess convergence and trigger early stopping, and to select the best-performing model, thereby limiting generalisation error associated with overfitting \citep{goodfellow2016deep}.
All post-training analysis is performed on a separate test set, distinct from both the training and validation points, that plays no role in optimisation, early stopping, or model selection.

The data-point sampling strategy is chosen to mimic observations from a physical, rather than a simulated, system.
We take $N_{\rm obs} = 2^n + 1$ measurements of $b(1,t)$ from the simulation at equal intervals across the training window, including both endpoints.
These intervals are multiples of the simulation time step, so no temporal interpolation is required.
From this set, $N_{\rm d}$ randomly chosen measurements form the training set and the remaining $N_{\rm d}^{\rm val} = N_{\rm obs} - N_{\rm d}$ form the validation set.
We take $N_{\rm d}$ to be $4N_{\rm obs}/5$ rounded to the nearest integer, giving approximately one validation point for every four training points.
The training data are therefore not uniformly distributed in time, but contain gaps where the validation points were removed.

Since the physics constraints are not tied to observations, the collocation points can be located at arbitrary coordinates within the domain.
We tested four approaches for placing the collocation points, focusing on schemes that give more uniform coverage of the domain than independent random draws.
These were Sobol sequences, Latin hypercube sampling, a regular grid shared by the $z^+$ and $z^-$ terms, and the same grid with the points of the two terms offset from each other by half a grid spacing in both $x$ and $t$.
We find that the specific sampling strategy has only a small effect on the reconstruction accuracy.
We therefore use Sobol sequences for the results presented in this paper, as these provide good coverage of the domain without imposing a regular grid structure that may introduce preferred directions and periodicities.
Training and validation points are drawn from independent Sobol sequences, separately for each term.
Since Sobol sequences achieve their best uniformity for sample counts that are powers of two, we use power-of-two values of $N_{z^\pm}$ and set $N_{z^\pm}^{\rm val} = N_{z^\pm}/4$, so that the validation sequence also contains a power-of-two number of points.
In all tests considered here, $N_{z^+} = N_{z^-}$.

The boundary-condition term in Eq.~\ref{eq:pinn_loss} is imposed at the domain boundary $x=1$, so the corresponding boundary points vary only in time.
As with the other collocation points, we draw the training and validation points from independent Sobol sequences, with $N_{\rm bc}$ a power of two and $N_{\rm bc}^{\rm val} = N_{\rm bc}/4$.

We choose the number of points in each loss term to be sufficiently large for the reconstruction accuracy to have converged, giving the values listed in Table~\ref{tab:fiducial_config}.
The convergence tests motivating these choices, and the choice of network size, are presented in \S\ref{sec:network_and_samples}.

Test points are sampled on a regular grid covering $(x,t) \in [0,1] \times [20,30]$ (including endpoints), with $257$ points in each of $x$ and $t$.
The temporal grid spacing is an integer multiple of the simulation time step, so the test data can be extracted directly from the simulation without temporal interpolation.
The simulation fields are directly sampled on this same spatial grid at every time step, so no spatial interpolation is required either.

\subsection{Optimiser}\label{subsec:optimiser}

We use the quasi-second-order SOAP optimiser \citep{vyas2025soap}, which has been shown to be effective for PINNs \citep{wang2025gradient, khodakarami2026spectral}. 
Our hyperparameters, listed in \ref{app:hyperparameters}, follow those used by \citet{wang2025gradient}, except that we use $(\beta_1, \beta_2) = (0.99, 0.99)$, which we found to improve accuracy, a constant learning rate, and update the preconditioner at every iteration.
Although all results presented here use full-batch training, SOAP is compatible with minibatching, making our validation more appropriate for the practical constraints of more demanding problems, where minibatching is required, than a second-order optimiser such as SS-Broyden \citep{kiyani2025optimizing}.

For the method comparisons in \S\ref{subsec:method_comparison}, we also train PINNs with the first-order Adam optimiser \citep{kingma2014adam}, which is the most common choice for PINNs, using the default \pytorch hyperparameters (\ref{app:hyperparameters}).

\subsection{Early stopping and best-model selection}\label{subsec:early_stopping}

At every training iteration, we evaluate validation counterparts of $\mathcal{L}_{\rm d}(\boldsymbol{\theta})$, $\mathcal{L}_{z^+}(\boldsymbol{\theta})$, $\mathcal{L}_{z^-}(\boldsymbol{\theta})$, and $\mathcal{L}_{\rm bc}(\boldsymbol{\theta})$ on each term's separate set of validation points.
Unlike the training loss, we set all pointwise weights $\lambda_i^j = 1$, recovering an unweighted mean squared error (MSE) for each term.
We then take the geometric mean of the four validation loss terms (three when the boundary-condition term is omitted, \S\ref{subsec:no_bc}), and use this to assess convergence and to select the best model.
The geometric mean is used since this prevents widely varying scales from allowing any single loss term to dominate the measure of convergence.

Early stopping is triggered when the lowest recorded geometric mean of validation losses has improved by a fraction less than $0.01$ over the preceding $5000$ iterations.
Early stopping and best-model selection are enabled only after iteration $10000$, allowing initial training transients to pass before assessing convergence.
Note that the loss-term and pointwise weights are still updated throughout this initial period.
Once early stopping has been triggered, the network parameters of the model with the lowest recorded geometric mean of validation losses are saved, and this model is used for the subsequent post-training analysis.

\subsection{Evaluation}\label{subsec:evaluation}

We train three independent repeats of every PINN.
For each repeat, the PINN is initialised with different weights and all training and validation point sets are resampled with different random seeds, including the validation points selected from the observations of $b_{\rm obs}$.

We assess reconstruction accuracy using two complementary measures of error: the local pointwise misfit and the global relative $L_2$ error.
The local pointwise misfit quantifies the absolute difference between the PINN reconstruction and reference solution at each test point $\mathbf{x}_i$,
\begin{equation}\label{eq:pointwise_error}
\varepsilon_\phi(\mathbf{x}_i) = \left| \phi_{\boldsymbol{\theta}}(\mathbf{x}_i) - \phi(\mathbf{x}_i) \right| \;,
\end{equation}
where $\phi \in \{u,b\}$ is the reference solution for either field and $\phi_{\boldsymbol{\theta}}$ is the corresponding PINN prediction.
This measure is used to examine the spatial and temporal distribution of the reconstruction error.

The global reconstruction accuracy is quantified using the relative $L_2$ error of the best-performing model on the grid of test points.
For each field, this is defined as
\begin{equation}\label{eq:rel_l2}
\mathcal{E}_\phi = \sqrt{\frac{
\sum_i \left[\phi_{\boldsymbol{\theta}}(\mathbf{x}_i) - \phi(\mathbf{x}_i)\right]^2
}{
\sum_i \left[\phi(\mathbf{x}_i)\right]^2
}}\;.
\end{equation}
We combine the errors of the $u$ and $b$ fields into a single measure,
\begin{equation}\label{eq:mean_error}
\mathcal{E} = \frac{1}{2}\left(\mathcal{E}_u + \mathcal{E}_b\right)\;,
\end{equation}
which we refer to as the ``mean relative $L_2$ error''.
This error is evaluated on the test points within the evaluation window, $t \in [21,29]$: removing a buffer of one Alfv\'en crossing time from each end of the training window $t \in [20,30]$ to avoid edge effects associated with the temporal domain boundaries, which we discuss in \S\ref{subsec:pinn_reconstruction}.

\section{Network size and numbers of training points}\label{sec:network_and_samples}

Here we present convergence tests that motivate our choices of network size and the number of training points used for each loss term.
In each test, we vary one of these quantities while keeping the others fixed at the values given in Table~\ref{tab:fiducial_config}, which constitute our final choices and are used throughout the other tests presented here.
These values are chosen to avoid any one of them playing a dominant role in limiting the accuracy of the reconstruction.
Because these choices are coupled, the final values were in practice determined iteratively.
All tests use the full PINN formulation of \S\ref{sec:PINNs}.

\begin{figure}
	\centering
{\includegraphics{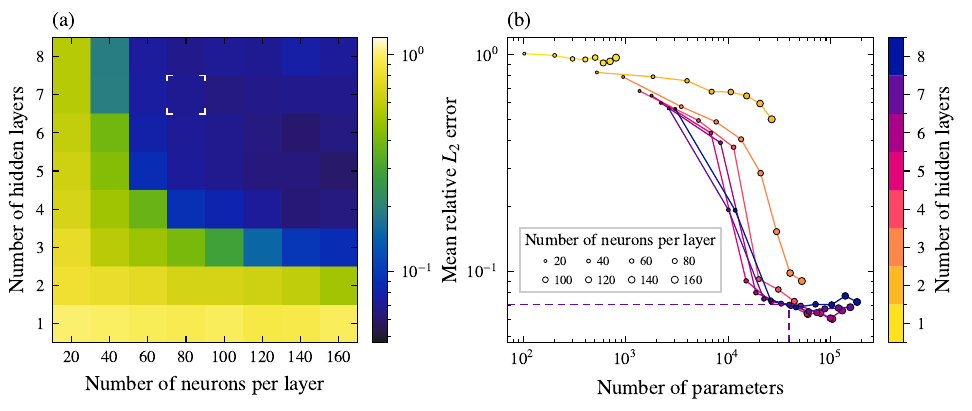}
}
    \vspace{-0.25in}
    \caption{Mean relative $L_2$ error of PINNs trained on the three-frequency fiducial system for networks with different numbers of hidden layers (depth) and neurons per hidden layer (width). Each value is the mean over three independent repeats. Panels show (a) error for each combination of depth and width, and (b) the same results plotted against the total number of network parameters, with colour indicating the number of hidden layers and marker size indicating the number of neurons per hidden layer. The selected architecture, with $7$ hidden layers of $80$ neurons, is marked by the corner brackets in (a) and by dashed lines in (b), drawn in the colour of its marker.
}
	\label{fig:test_error_combined}
\end{figure}

We first vary the network size, considering for simplicity only networks in which all hidden layers have the same width.
We use the three-frequency system as the basis for this test, as it is the most complex of the fiducial systems and therefore provides the most demanding test of the network's representational capacity.
Fig.~\ref{fig:test_error_combined} shows the mean relative $L_2$ error (Eq.~\ref{eq:mean_error}), averaged over three repeats, for networks with $1$--$8$ hidden layers and $20$--$160$ neurons per hidden layer, in steps of $20$.
The grid in Fig.~\ref{fig:test_error_combined}(a) shows a general decrease in error with increasing network size, both through increasing the number of hidden layers and the number of neurons per layer.

Each layer $\ell$ contributes an $N_\ell \times N_{\ell-1}$ weight matrix and a bias vector of length $N_\ell$, giving the total number of network parameters
\begin{equation}\label{eq:n_params}
    N_{\rm params} = \sum_{\ell=1}^{L} \left(N_{\ell-1} + 1\right) N_\ell \;,
\end{equation}
where our PINNs have $N_0 \equiv N_{\rm in} = 2$ and $N_L \equiv N_{\rm out} = 2$, and $N_1, \dots, N_{L-1}$ are the hidden-layer widths.
In Fig.~\ref{fig:test_error_combined}(b), the same results as in (a) are plotted against the total number of network parameters.
There is a sharp increase in accuracy for sufficiently deep networks at ${\sim}10^4$ parameters, with the results for these networks appearing to collapse onto a common curve.

To select a network size to use for subsequent tests, we define a reference error as the mean of the ten lowest repeat-averaged errors across all tested network sizes (the mean $y$-axis value of the ten lowest points in Fig.~\ref{fig:test_error_combined}(b)).
This is less sensitive to noise than using the single lowest point.
We then choose the architecture with the fewest parameters whose repeat-averaged error lies within $10\%$ of this reference.
For the network sizes tested, this gives a network with $7$ hidden layers of $80$ neurons each, which achieves close to the best accuracy while avoiding the additional training cost of the largest networks.

\begin{figure}
	\centering
{\includegraphics{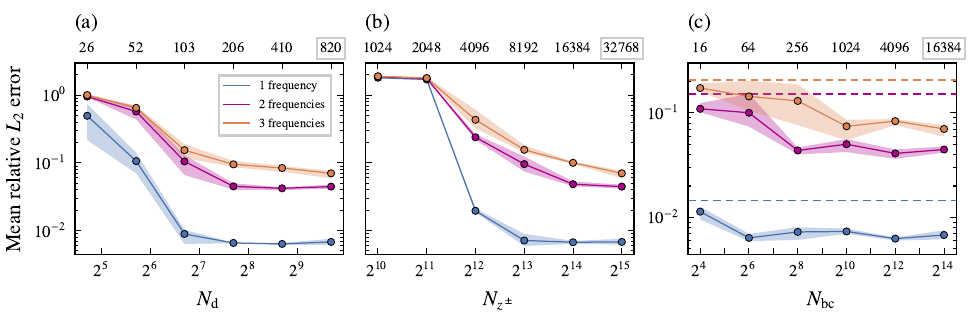}
}
    \vspace{-0.25in}
    \caption{Mean relative $L_2$ error plotted against the number of training points used for (a) the data loss; (b) each physics loss, with both $z^\pm$ varied together; and (c) the boundary-condition loss, for PINNs trained on the fiducial systems forced with one, two, and three frequencies. In each panel, the numbers of points in the other loss terms are held at the values given in Table~\ref{tab:fiducial_config}. Each plotted marker corresponds to the mean of three repeats, and the shaded regions span the minimum to maximum values across the repeats. The top axis of each panel gives the exact number of training points at each tested value, with the grey outline indicating the selected values. The horizontal dashed lines in (c) show the mean over three repeats of PINNs trained without the boundary-condition loss term ($N_{\rm bc} = 0$), for each of the three systems.
}
	\label{fig:sampling}
\end{figure}

We next vary the number of training points in each loss term, keeping $N_j^{\rm val} = N_j/4$ (approximately, in the case of the data loss) and $N_{z^+} = N_{z^-}$.
Results for all three fiducial systems are shown in Fig.~\ref{fig:sampling}.
The convergence is well behaved for all three systems.
The error generally decreases with increasing numbers of training points, and the spread between repeats is small compared with the differences between the individual systems.
The one-frequency system reaches errors approximately an order of magnitude lower than the two- and three-frequency systems, with the three-frequency system giving slightly higher errors than the two-frequency system.
This reflects the frequency content of the systems: each additional forcing frequency introduces finer-scale structure, making the resulting dynamics harder for the network to learn accurately.
Since the training window spans ten Alfv\'en crossing times, the tested values of $N_{\rm d}$ correspond to between $2.6$ and $82$ data points per crossing time.
The number of data points needed to reach an error of ${\sim}0.1$ increases with the number of forcing frequencies, from ${\sim}5$ per crossing time for the one-frequency system to ${\sim}20$ for the three-frequency system.

The dependence on the number of boundary-condition points is weaker, as shown in Fig.~\ref{fig:sampling}(c).
With $2^{14}$ training points, including the boundary-condition term reduces the error by a factor of approximately 2--4 relative to PINNs trained without it (dashed lines).
Although this difference is modest compared with the variation in (a) and (b), omitting the boundary-condition term produces errors with a distinct spatial and temporal structure, which are examined in \S\ref{subsec:no_bc}.

For all other PINNs presented here, we use the values in Table~\ref{tab:fiducial_config}. The data points are sampled from $N_{\rm obs} = 1025 = 2^{10}+1$ observations, and the other training sets are powers of two: $N_{z^\pm} = 2^{15}$ and $N_{\rm bc} = 2^{14}$.

\section{Results and discussion}\label{sec:results}

\begin{figure}
	\centering
{\includegraphics{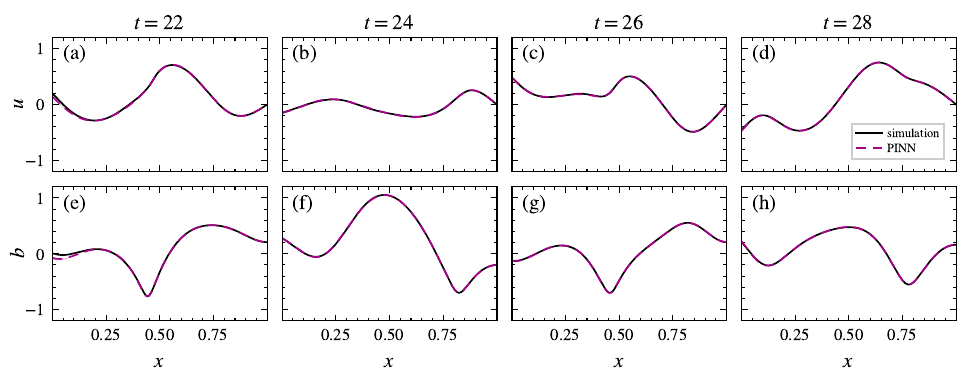}
}
    \vspace{-0.25in}
	\caption{Underlying simulation (solid) and PINN reconstruction from boundary measurements (dashed) for (a--d) velocity, $u$, and (e--h) magnetic field perturbation, $b$, in the two-frequency fiducial system at $t = 22$, $24$, $26$, and $28$. Dashed lines show one of three independent repeats. Shaded regions span the minimum to maximum values across the three repeats at each position; they are generally narrower than the plotted linewidth and are therefore largely hidden.}
	\label{fig:field_snapshots_freq02_repeat01}
\end{figure}

Using the network size and numbers of training points chosen in \S\ref{sec:network_and_samples}, we now assess how well PINNs are able to reconstruct our one-dimensional magnetohydrodynamic system.
In \S\ref{subsec:pinn_reconstruction}, we apply the PINN formulation described in \S\ref{sec:PINNs} to the three fiducial systems introduced in \S\ref{subsec:fiducial}, forced with one, two, and three frequencies.
In \S\ref{subsec:method_comparison}, we then examine the effect of each modification to the basic PINN formulation.
In \S\ref{subsec:no_bc}, we test reconstruction under fewer constraints by omitting the loss term for the Neumann boundary condition on the magnetic field perturbation, the only constraint in the loss besides the observations and the governing equations.

\subsection{PINN reconstruction of one-dimensional MHD system}\label{subsec:pinn_reconstruction}

We first apply the PINN formulation outlined in \S\ref{sec:PINNs} to the fiducial systems of \S\ref{subsec:fiducial}, forced with one, two, and three frequencies.
Our PINN formulation uses a SIREN network with 7 hidden layers of 80 neurons, trained with the SOAP optimiser, dynamic loss-term weights, and residual-based attention.
Boundary measurements of the magnetic field perturbation, $b_{\rm obs}(t)$, are the only time-dependent information provided to the networks for training.
The training dynamics of these PINNs are examined in more detail in \ref{app:training}.

A comparison between the simulated and PINN-reconstructed $u$ and $b$ fields of the two-frequency system is shown in Fig.~\ref{fig:field_snapshots_freq02_repeat01}.
The reconstruction closely matches the simulated fields across the domain, capturing both sharply varying and more subtle features.
We carry out three independent repeats, using differently sampled datasets and differently initialised network parameters, all of which show similarly close agreement with the simulation.

\begin{figure}[!t]
	\centering
{\includegraphics{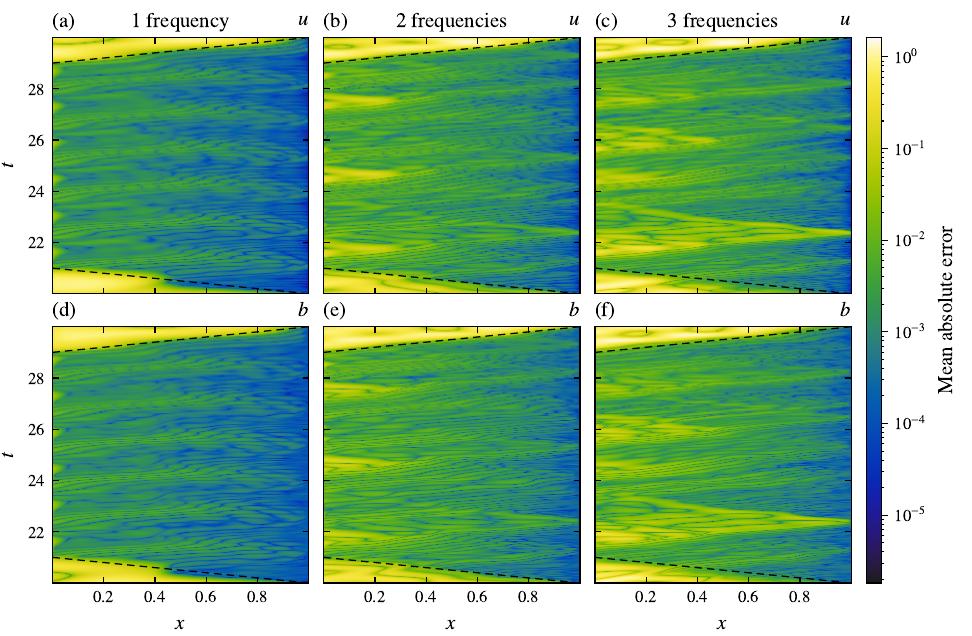}
}
    \vspace{-0.25in}
	\caption{Pointwise absolute error of the PINN reconstruction of (a--c) velocity, $u$, and (d--f) magnetic field perturbation, $b$, over the training window $t \in [20, 30]$, for the fiducial systems forced with (a, d) one, (b, e) two, and (c, f) three frequencies. Error values of each pixel are averaged over three independent repeats. The dashed lines have slopes $\pm1$, corresponding to the Alfv\'en speed, and intersect the observation point $x = 1$ at the lower and upper edges of the training window.}
	\label{fig:test_error_grid_mean}
\end{figure}

In Fig.~\ref{fig:test_error_grid_mean}, the pointwise error (Eq.~\ref{eq:pointwise_error}) is shown on the grid of test data points for all three frequency cases, with the value of each pixel averaged over three PINN repeats.
The error is low throughout most of the domain, typically orders of magnitude below the $\mathcal{O}(1)$ amplitude of the fields.
It is higher for systems with more forcing frequencies and lowest near $x = 1$, where $b$ is directly observed and the boundary conditions are imposed.
Further from $x = 1$, the reconstruction relies increasingly on the physics-loss constraints to carry information from the boundary observations into the interior, and so becomes more sensitive to how well these constraints are satisfied.
Because the PINN is trained on the full domain at once, a localised error anywhere can in principle influence the fit everywhere through the shared network parameters.
However, when the physics-loss constraints are generally well satisfied, localised errors can propagate through the system by the governing dynamics, just as a physical perturbation would, but can affect points both forward and backward in time.
To first order, information propagates through the system at the Alfv\'en speed ($1$ in our units).
Consequently, localised errors near $x=1$ spread into the interior both forward and backward in time, producing triangular error structures bounded by lines of slope $\pm1$.

The largest errors occur in wedge-shaped regions near the temporal edges of the training window, bounded approximately by the dashed lines in Fig.~\ref{fig:test_error_grid_mean}, whose slopes correspond to the Alfv\'en speed; the bound is only approximate because the nonlinear terms modify how information is transported through the system.
Near the start of the window, reconstructing the system would require information about the leftward-propagating $z^+$ that would have had to be measured (at $x=1$) earlier than the available observations, while near the end of the window, the rightward-propagating $z^-$ would reach the boundary later than the end of the training window.
Therefore, these high-error regions arise from the limited temporal extent of the observations combining with the lack of additional information to constrain the solution outside this temporal range, since the PINN is not, for example, given a prior functional form for the forcing that could be extrapolated beyond the training window.
This edge effect motivates the evaluation window, $t \in [21, 29]$, used in our global measure of mean relative $L_2$ error (\S\ref{subsec:evaluation}).
Additional tests using different training-window sizes show that reconstruction accuracy away from the window edges is not sensitive to the window size, while the high-error edge effects shift with the window boundaries.

For each of the three fiducial systems, we trained three independent repeats, each running on a single NVIDIA H200 NVL GPU.
Early stopping was triggered after ${\sim}35000$--$51000$ iterations for the one-frequency system, ${\sim}55000$--$63000$ for the two-frequency system, and ${\sim}46000$--$74000$ for the three-frequency system, corresponding to wall-clock training times of $0.6$--$0.9$\,h, $1.0$--$1.1$\,h, and $0.8$--$1.3$\,h per PINN, respectively.

\subsection{PINN method comparison}\label{subsec:method_comparison}

\begin{figure}[!t]
	\centering
{\hspace*{0.95in}\includegraphics{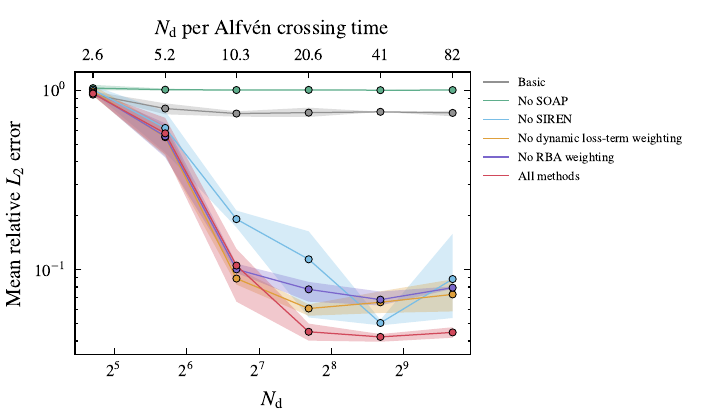}
}
    \vspace{-0.10in}
	\caption{Mean relative $L_2$ error as a function of the number of data-loss training points for the two-frequency fiducial system, for PINNs trained using different combinations of methods.
Results are shown for the Basic formulation, the full formulation described in \S\ref{sec:PINNs} (All methods), and the full formulation with each method removed individually.
Each point corresponds to the mean over three independent repeats, with shaded regions spanning the minimum to maximum values. The top axis gives the $N_{\rm d}$ per Alfv\'en crossing time of each point, with the training window spanning ten crossing times.}
	\label{fig:test_error_vs_data_methods}
\end{figure}

The PINNs in \S\ref{subsec:pinn_reconstruction} combine four modifications to the original PINN formulation of \citet{raissi2019physics}, which set out a basic framework for implementation. These are: the SIREN architecture, the SOAP optimiser, dynamic loss-term weighting, and residual-based attention.
We refer to the formulation without any of these modifications as the ``Basic'' formulation: an MLP with hyperbolic tangent activation functions, trained with the Adam optimiser, with all loss-term and pointwise weights fixed to $1$.
We compare this with the full formulation of \S\ref{sec:PINNs} (``All methods'') and with four cases in which a single method is reverted to its Basic counterpart (``No \textit{method}'').
All other settings are unchanged and are listed in \ref{app:hyperparameters}.

The mean relative $L_2$ error for each formulation is shown in Fig.~\ref{fig:test_error_vs_data_methods} for the two-frequency system, as a function of the number of data points used for training and with all other parameters and choices kept the same.
Since the training window spans ten Alfv\'en crossing times, the tested values of $N_{\rm d}$ correspond to between $2.6$ and $82$ data points per crossing time, as shown on the top axis.
With the fewest data points tested, all formulations give $\mathcal{O}(1)$ errors.
As the number of data points increases, the error of the full formulation falls before levelling off at ${\sim}20$ data points per crossing time, beyond which it gives the lowest errors of the schemes tested.
In contrast, the errors of the Basic and No SOAP formulations remain of $\mathcal{O}(1)$ regardless of the number of data points; the No SOAP formulation converges to a trivial solution, with $u \approx 0$ and $b \approx 0$ throughout the interior, and the Basic formulation gives a solution close to this trivial one, with some perturbations.
Removing either weighting scheme produces an error convergence behaviour similar to that of the full formulation, but the error levels off at a higher value.
Without SIREN, the error decreases less rapidly and less consistently, with greater variation between repeats.
In additional tests, we found the behaviour of the No SIREN formulation to be sensitive to the network size, which was chosen in \S\ref{sec:network_and_samples} for the SIREN architecture, and so the results shown here may not reflect its convergence with a network size chosen specifically with this architecture in mind.

\subsection{Effect of omitting the boundary-condition loss term}\label{subsec:no_bc}

\begin{figure}[!t]
	\centering
{\includegraphics{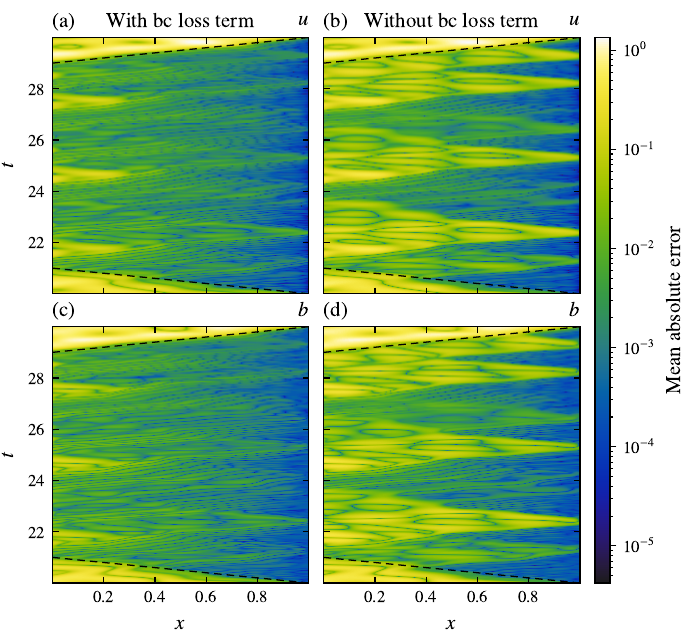}
}
    \vspace{-0.15in}
	\caption{Pointwise absolute error of the PINN reconstruction of (a, b) velocity, $u$, and (c, d) magnetic field perturbation, $b$, for the two-frequency fiducial system, for PINNs trained (a, c) with and (b, d) without the boundary-condition loss term. Error values are averaged over three independent repeats. Dashed lines are as in Fig.~\ref{fig:test_error_grid_mean}.}
	\label{fig:nobc}
\end{figure}

In the PINN formulation presented in \S\ref{sec:PINNs}, the Neumann condition $\partial_x b(1,t) = 0$ is imposed through the boundary-condition loss term.
Like the hard constraint $u(1,t) = 0$, this condition is not a measurement, but instead represents prior knowledge of the system's behaviour at the boundary and provides additional information that constrains the PINN during training.
In physical inverse problems, such prior knowledge may be uncertain or unavailable, and its absence may leave the problem under-constrained, so that the observations and governing equations alone do not uniquely determine the system state.
In our formulation, the boundary conditions at $x=0$, including the forcing, are withheld from the PINN, and so the Neumann condition on $b$ is the only boundary condition imposed through a loss term.
To assess how the reconstruction is affected by removing this information, we omit the corresponding term from Eq.~\ref{eq:pinn_loss}, leaving the data and physics terms and all other settings unchanged.

In Fig.~\ref{fig:nobc}, we compare the pointwise error of PINNs trained with and without the boundary-condition term for the two-frequency system.
With the loss term included, the figure panels are the same as those shown in Fig.~\ref{fig:test_error_grid_mean}(b) and (e).
Without it, the error is considerably larger over much of the interior, and the triangular error structures described in \S\ref{subsec:pinn_reconstruction}, which extend from the $x=1$ boundary both forward and backward in time, are much stronger.

The high-error structures are also consistent across repeats, with similar structures appearing at the same locations in each PINN reconstruction.
Their reproducibility suggests that they are a systematic feature of the reconstruction rather than a consequence of a particular training realisation.
Thus, the absence of the additional constraint does not appear to result in PINN repeats sampling different possible solutions that become consistent with the remaining constraints when the system is under-constrained, with the underlying solution recovered in some cases.
Instead, the PINNs consistently converge towards solutions exhibiting these specific error structures.
Additional tests with different numbers of data points suggest that the reconstruction error without the boundary-condition term has converged with respect to the number of data points.

The data and physics loss terms converge to similar values whether or not the boundary-condition term is included in training, despite the larger reconstruction error without it.
Thus, the training and validation loss values alone do not provide a clear indication of reconstruction quality.

\section{Conclusions}\label{sec:conclusions}

We have investigated the use of PINNs for reconstructing interior flow from boundary magnetic-field measurements in a one-dimensional MHD system.
Trained using observations of the magnetic field at a single boundary together with the governing equations and boundary conditions at $x=1$, the PINN reconstructs both the velocity and magnetic fields throughout the domain, with reconstruction errors typically orders of magnitude lower than the field amplitudes.
The only time-dependent information provided to the network is the magnetic field at the boundary.
Accurate reconstructions are obtained for systems with one, two, and three forcing frequencies, although the reconstruction error increases with the dynamical complexity of the system.
The density of boundary observations needed to drop below a mean relative $L_2$ error of ${\sim}0.1$ increases from ${\sim}5$ training observations per Alfv\'en crossing time for the one-frequency system to ${\sim}20$ for the three-frequency system.

The accuracy of the reconstruction depends on the combination of methods used in the PINN formulation.
The full combination of methods considered here, which consists of a SIREN architecture, the SOAP optimiser, dynamic loss-term weighting, and residual-based attention, gives the lowest reconstruction errors.
In contrast, the basic PINN formulation, and the formulation that uses the Adam optimiser instead of SOAP, fail to recover the interior dynamics.
Removing any of the other methods also degrades the reconstruction or increases the amount of data required to achieve comparable accuracy.
These results indicate that each modification contributes a cumulative improvement to the reconstruction, consistent with the different methods addressing distinct difficulties in PINN training.

Removing the loss term based on the Neumann condition on the magnetic field increases the reconstruction error and produces structures of error that originate at the observation boundary and are consistent across independent repeats.
The reproducibility of these error features demonstrates that the reliable convergence of PINNs to a specific solution does not by itself indicate that the underlying system has been recovered.
For under-constrained physical inverse problems, this highlights the difficulty of assessing reconstruction accuracy from either the final solution or the training behaviour when an incorrect solution can nevertheless exhibit reproducible convergence.

The system considered here is deliberately idealised.
The observations are noise-free, the dimensionless numbers and governing equations are known, and the dynamics are one-dimensional.
The effects of observational noise, uncertain or unknown system parameters, and higher-dimensional dynamics therefore remain to be established.
These extensions, which are to be explored in future work, are important for determining how the approach performs under the conditions encountered in physical MHD inverse problems.
In addition, since the boundary conditions at $x=0$, including the forcing of the system, are not provided to the PINN, the governing equations and boundary conditions it is given do not form a closed problem, and the reconstruction cannot be extended beyond the training window as a prediction.
Providing these boundary conditions would close the problem, but for our one-dimensional system would also make it considerably easier, since the forcing would no longer need to be inferred from the observations.

Overall, the results presented here demonstrate that PINNs can be used to reconstruct an evolving interior MHD system from magnetic-field measurements taken only at the system boundary.
We show that the quality of the reconstruction is sensitive to the combination of methods used and the choices made in the formulation and training of the PINN.
The one-dimensional MHD testbed presented here provides a controlled setting for validating and benchmarking data assimilation approaches before extending them to physical inverse problems in which the interior state cannot be observed directly.

\section*{Acknowledgements}
\noindent
We thank Chris Finlay and Nicolas Gillet for valuable discussions.
This work was supported by Natural Environment Research Council (NERC) Grant UKRI1268 and by ESA in the framework of EO Science for Society through contract 4000148713/25/NL/FFi (4D Earth Core+). PWL was additionally supported by Swarm DISC activities funded by ESA contract No. 4000109587, and JL by the Macau Foundation.
This work was undertaken on Calder, part of the High Performance Computing facilities at the University of Leeds, UK.

\section*{Data availability}
\noindent
The code used to produce all results presented here is publicly available at \coderepo\ (v1.0.0).
This includes the scripts for running and sampling the simulations, and the problem-specific set-up for training and analysing the PINNs.
The training itself is carried out using \pinninit (v0.1.0), our general-purpose PINN framework, which is available separately at \pinninitrepo.
The trained PINN models presented here will be made available on publication.
All simulation data and trained models can also be regenerated from the code.

\section*{Declaration of generative AI in the manuscript preparation process}
\noindent
The authors used Claude (Anthropic) to assist with editing the manuscript. The authors reviewed and edited the content as needed and take full responsibility for the content of the published article.

\appendix
\counterwithin*{figure}{section}
\counterwithin*{table}{section}

\section{Derivation of the dimensionless governing equations}\label{app:derivation}

Here we derive Eq.~\ref{eq:elsasser_eqns} from Eqs.~\ref{eq:thomas_eqns_u} and~\ref{eq:thomas_eqns_b}.
First, substituting $B = U_A + b$ into Eqs.~\ref{eq:thomas_eqns_u} and~\ref{eq:thomas_eqns_b}, where $U_A$ is the constant background magnetic field, gives
\begin{align}
    \frac{\partial u}{\partial t} - U_A \frac{\partial b}{\partial x} + u \frac{\partial u}{\partial x} - b \frac{\partial b}{\partial x} &= \nu \frac{\partial^2 u}{\partial x^2} \;, \\
    \frac{\partial b}{\partial t} - U_A \frac{\partial u}{\partial x} + u \frac{\partial b}{\partial x} - b \frac{\partial u}{\partial x} &= \eta \frac{\partial^2 b}{\partial x^2} \;.
\end{align}
We non-dimensionalise by writing $x = L\tilde{x}$, $t = T\tilde{t}$ with the Alfv\'en crossing time $T \equiv L/U_A$, and $u = U\tilde{u}$, $b = U\tilde{b}$, where tildes denote dimensionless quantities.
This gives
\begin{align}
    \frac{U U_A}{L} \left( \frac{\partial \tilde{u}}{\partial \tilde{t}} - \frac{\partial \tilde{b}}{\partial \tilde{x}} \right) + \frac{U^2}{L} \left( \tilde{u} \frac{\partial \tilde{u}}{\partial \tilde{x}} - \tilde{b} \frac{\partial \tilde{b}}{\partial \tilde{x}} \right) &= \frac{\nu U}{L^2} \frac{\partial^2 \tilde{u}}{\partial \tilde{x}^2} \;, \\
    \frac{U U_A}{L} \left( \frac{\partial \tilde{b}}{\partial \tilde{t}} - \frac{\partial \tilde{u}}{\partial \tilde{x}} \right) + \frac{U^2}{L} \left( \tilde{u} \frac{\partial \tilde{b}}{\partial \tilde{x}} - \tilde{b} \frac{\partial \tilde{u}}{\partial \tilde{x}} \right) &= \frac{\eta U}{L^2} \frac{\partial^2 \tilde{b}}{\partial \tilde{x}^2} \;.
\end{align}
Dividing both equations by $U U_A/L$, dropping the tildes for convenience, and using $U/U_A \equiv M_A$, $\nu/(U_A L) \equiv M_A/Re$, and $\eta/(U_A L) \equiv M_A/R_m$ gives
\begin{align}
    \frac{\partial u}{\partial t} - \frac{\partial b}{\partial x} + M_A \left( u \frac{\partial u}{\partial x} - b \frac{\partial b}{\partial x} \right) &= \frac{M_A}{Re} \frac{\partial^2 u}{\partial x^2} \;, \label{eq:app_u}\\
    \frac{\partial b}{\partial t} - \frac{\partial u}{\partial x} + M_A \left( u \frac{\partial b}{\partial x} - b \frac{\partial u}{\partial x} \right) &= \frac{M_A}{R_m} \frac{\partial^2 b}{\partial x^2} \;. \label{eq:app_b}
\end{align}
Adding and subtracting Eqs.~\ref{eq:app_u} and~\ref{eq:app_b} gives, in terms of the Els\"asser variables $z^\pm \equiv u \pm b$,
\begin{equation*}
    \frac{\partial z^\pm}{\partial t} \mp \frac{\partial z^\pm}{\partial x} + M_A \, z^\mp \frac{\partial z^\pm}{\partial x} = \nu^+ \frac{\partial^2 z^\pm}{\partial x^2} + \nu^- \frac{\partial^2 z^\mp}{\partial x^2} \;,
\end{equation*}
with $\nu^\pm \equiv M_A \left(1/Re \pm 1/R_m\right)/2$, recovering Eq.~\ref{eq:elsasser_eqns}.

\section{PINN hyperparameters}\label{app:hyperparameters}

The hyperparameters used for the PINNs presented here are listed in Table~\ref{tab:hyperparameters}.
All training is full-batch, so each iteration corresponds to one epoch.
The settings used for the alternative methods in the comparisons of \S\ref{subsec:method_comparison} are listed in Table~\ref{tab:hyperparameters_comparison}.

\begin{table}
\centering
\caption{PINN hyperparameters.}\label{tab:hyperparameters}
\vspace{0.10in}
\begin{tabular}{lr}
\toprule
Quantity & Value \\
\midrule
\multicolumn{2}{l}{\textit{Network architecture} (\S\ref{subsec:architecture})} \\
Architecture                                        & SIREN \\
Input rescaling                                     & $x, t \mapsto [-1, 1]$ \\
Activation function                                 & $\sin$ \\
$\omega_0$                                          & 30 \\
Floating-point precision                            & 64-bit \\
\midrule
\multicolumn{2}{l}{\textit{SOAP optimiser} (\S\ref{subsec:optimiser})} \\
Reference implementation                            &\href{https://github.com/nikhilvyas/SOAP}{\texttt{github.com/nikhilvyas/SOAP}} \\
Learning rate                                       & $10^{-3}$ \\
$(\beta_1, \beta_2)$                                & $(0.99, 0.99)$ \\
$\beta_{\rm shampoo}$                               & $\beta_2$ \\
$\epsilon$                                          & $10^{-8}$ \\
Weight decay                                        & 0 \\
Precondition frequency                              & 1 \\
Maximum preconditioner dimension                    & 10000 \\
Merge preconditioner dimensions                     & False \\
Precondition 1D gradients                           & False \\
Normalise gradients                                 & False \\
Correct bias                                        & True \\
\midrule
\multicolumn{2}{l}{\textit{Dynamic loss-term weighting} (\S\ref{subsec:loss})} \\
Initial loss-term weights                           & 1 \\
$\alpha$                                            & 0.9 \\
Update frequency                                    & 1000 \\
\midrule
\multicolumn{2}{l}{\textit{Residual-based attention} (\S\ref{subsec:loss})} \\
Initial pointwise weights                           & 1 \\
$\gamma$                                            & 0.999 \\
$\eta^*$                                            & 0.01 \\
Update frequency                                    & 1 \\
\midrule
\multicolumn{2}{l}{\textit{Early stopping and best-model selection} (\S\ref{subsec:early_stopping})} \\
Monitored quantity                                  & Geometric mean of validation MSEs \\
Iterations before monitoring                        & 10000 \\
Patience                                            & 5000 \\
Fractional tolerance                                & 0.01 \\
\bottomrule
\end{tabular}
\end{table}

\begin{table}
\centering
\caption{Alternative settings used in the method comparisons.}\label{tab:hyperparameters_comparison}
\vspace{0.10in}
\begin{tabular}{lr}
\toprule
Quantity & Value \\
\midrule
Activation function                                 & $\tanh$ \\
Optimiser                                           & Adam \\
Learning rate                                       & $10^{-3}$ \\
$(\beta_1, \beta_2)$                                & $(0.9, 0.999)$ \\
$\epsilon$                                          & $10^{-8}$ \\
Weight decay                                        & 0 \\
Constant loss-term weight value                     & 1 \\
Constant pointwise weight value                     & 1 \\
\bottomrule
\end{tabular}
\end{table}

\section{Training dynamics}\label{app:training}

Here we examine the training behaviour of the PINNs presented in \S\ref{subsec:pinn_reconstruction}.
We consider the evolution of each loss term in Eq.~\ref{eq:pinn_loss} evaluated on its training and validation points, together with the evolution of the loss-term weights and the distribution of RBA weights.
These methods are described in \S\ref{subsec:loss}.
We show these quantities as a function of training epoch; since no minibatching is used in this study, each epoch corresponds to one optimiser iteration that updates the network parameters using the full training datasets.

\begin{figure}
	\centering
{\includegraphics{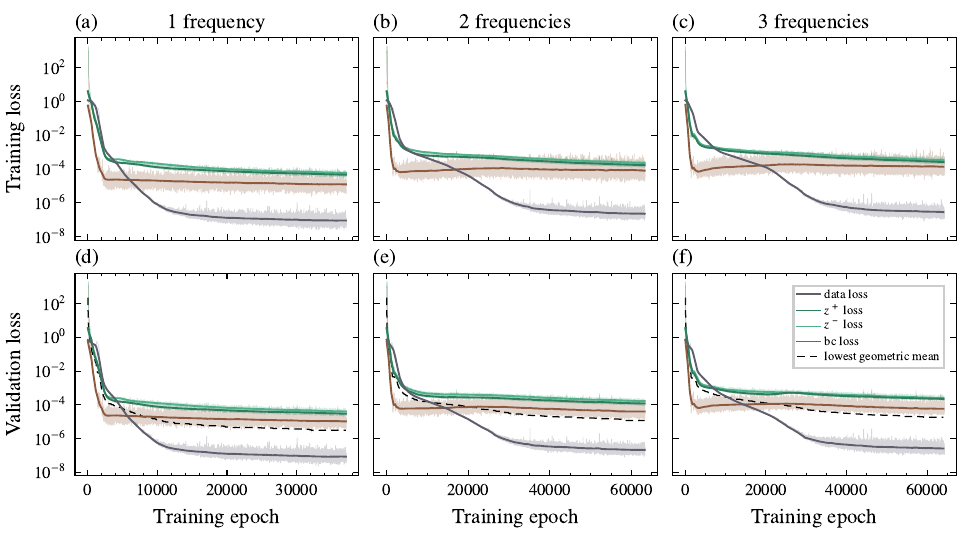}
}
    \vspace{-0.25in}
    \caption{Loss terms evaluated on their (a--c) training and (d--f) validation points as a function of training epoch, for one repeat of the PINNs trained on the fiducial systems forced with one, two, and three frequencies. Faint lines show the value at every epoch, and solid lines show a rolling geometric mean over a centred window of 1000 epochs. The dashed line shows the lowest geometric mean of the four validation losses recorded up to each epoch. Training losses are RBA-weighted mean squared residuals, validation losses are unweighted, and neither includes the loss-term weights. Training losses are evaluated before each epoch's parameter update and validation losses after it, and both are plotted against the number of parameter updates completed at the time of evaluation.}
	\label{fig:losses_vs_iteration}
\end{figure}

The evolution of each loss term during training, evaluated on the corresponding training and validation points and without the loss-term weights $w_j$, is shown in Fig.~\ref{fig:losses_vs_iteration}.
All four loss terms decrease by several orders of magnitude within the first few thousand iterations, after which they settle into approximately steady ranges with fluctuations on the frequency-scale of individual iterations.
The $z^+$ and $z^-$ loss terms closely follow each other throughout training, while the data loss falls lower than the other terms and continues to decrease after the other loss terms have largely stabilised.
There are slight variations between the cases with different numbers of forcing frequencies, but the overall trends are consistent across the tests.
For every loss term, the validation loss closely follows its training counterpart and does not increase as training continues, consistent with the PINNs not overfitting their training points.
The lowest recorded geometric mean of the validation loss terms decreases steadily before flattening towards the end of training, at which point early stopping is triggered.

\begin{figure}
	\centering
{\includegraphics{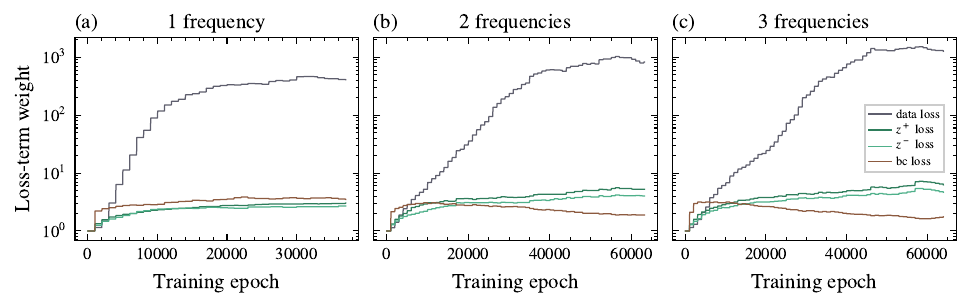}
}
    \vspace{-0.25in}
	\caption{Evolution of the four loss-term weights for PINNs trained to reconstruct the fiducial systems forced with one, two, and three frequencies. The weights are updated every $1000$ iterations, giving the stepped profiles.}
	\label{fig:w23_weights_vs_iteration}
\end{figure}

The evolution of the loss-term weights, $w_j$ (Eq.~\ref{eq:wang_23_w}), is shown in Fig.~\ref{fig:w23_weights_vs_iteration}.
While the physics and boundary-condition weights remain low and at a similar scale, the data-loss weight increases steadily to values of order $10^2$--$10^3$ before levelling off.
The loss-term weights are adaptively updated to rescale the individual loss terms so that their gradient norms with respect to the network parameters, and hence their contributions to the parameter updates, remain comparable throughout training.
As shown in Fig.~\ref{fig:losses_vs_iteration}, the data loss continues to decrease after the other loss terms have stabilised.
The corresponding increase in the data-loss weight compensates for the decreasing gradient of this loss term as the data loss decreases, keeping the data term influential in the parameter updates, consistent with its continued decrease.

\begin{figure}
	\centering
{\includegraphics{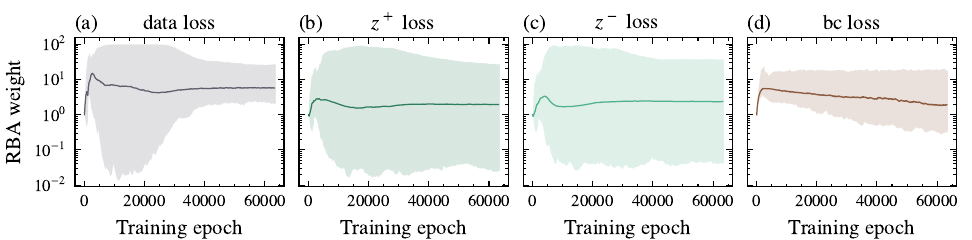}
}
    \vspace{-0.25in}
	\caption{Distribution of residual-based attention weights as a function of training epoch for the (a) data, (b) $z^+$, (c) $z^-$, and (d) boundary-condition loss terms, for a PINN trained on the two-frequency fiducial system. Solid lines show the median weight across the training points for each loss term, and shaded regions span the minimum to maximum values.
}
	\label{fig:rba_weights_vs_iteration_freq02_repeat01}
\end{figure}

The evolution of the distribution of RBA weights, $\lambda_i^j$ (Eq.~\ref{eq:rba_lambda}), for a PINN trained on the two-frequency fiducial system is shown in Fig.~\ref{fig:rba_weights_vs_iteration_freq02_repeat01}.
After some transient behaviour early in training, each distribution settles into a dynamically steady state with approximately consistent median, minimum, and maximum values.
The weights within each term span two to four orders of magnitude throughout training.
These weights cannot exceed a value of $100$ \citep{anagnostopoulos2024residual}, and the maximum weights do not consistently saturate this limit, suggesting that no single point retains the largest residual for long.
Because the averages in Eq.~\ref{eq:pinn_loss} are normalised by $\sum_i \lambda_i^j$, only the relative values of $\lambda_i^j$ within each term affect training.

\bibliographystyle{elsarticle-harv} 
\bibliography{bibliography.bib}

\end{document}